\documentclass[11pt]{amsart}
\usepackage{wrapfig}
\usepackage[english]{babel}
\usepackage{physics}
\usepackage{braket}
\usepackage{dsfont}
\usepackage{xcolor}
\usepackage{graphicx}
\usepackage{tikz}
\usetikzlibrary{matrix}
\usepackage{color,soul}
\usepackage[utf8]{inputenc}
\usepackage{mathtools}

\RequirePackage[framemethod=default]{mdframed}

\usepackage[margin=1in]{geometry}
\usepackage[pagebackref, colorlinks = true, linkcolor = blue, urlcolor  = blue, citecolor = red]{hyperref}

\renewcommand{\epsilon}{\varepsilon}

\def\rh{\rho_{AB}}
\def\ra{\rho_{A}}
\def\rb{\rho_{B}}
\def\sh{\sigma_{AB}}
\def\sa{\sigma_{A}}
\def\sb{\sigma_{B}}

\def\hs{\mathcal{H}}

\newcommand{\identity}{\ensuremath{\mathds{1}}}

\newcommand{\ent}[2]{D ( #1 || #2 )}
\newcommand{\bs}[2]{\widehat{D} ( #1 || #2 )}
\newcommand{\bsA}[3]{\widehat{D}_{#1} ( #2 || #3 )}
\newcommand{\entA}[3]{D_{#1}(#2||#3)}

\definecolor{grey}{RGB}{220,220,220}

\newmdenv[skipabove=7pt,
skipbelow=7pt,
backgroundcolor=grey!20,
innerleftmargin=5pt,
innerrightmargin=5pt,
innertopmargin=5pt,
leftmargin=0cm,
rightmargin=0cm,
innerbottommargin=5pt,
linewidth=1pt]{sBox}

\newmdenv[skipabove=7pt,
skipbelow=7pt,
backgroundcolor=grey!40,
innerleftmargin=5pt,
innerrightmargin=5pt,
innertopmargin=5pt,
leftmargin=0cm,
rightmargin=0cm,
innerbottommargin=5pt,
linewidth=1pt]{tBox}

\newtheorem{thm}{Theorem}[section]
\newtheorem*{thm*}{Theorem}

\newtheorem{defi}[thm]{Definition}

\newtheorem{remark}[thm]{Remark}

\newtheorem{stp}{Step}
\newtheorem*{stp2}{Step 3'}

\newenvironment{theo}{\begin{tBox}\begin{thm}}{\end{thm}\end{tBox}}
\newenvironment{step}{\begin{sBox}\begin{stp}}{\end{stp}\end{sBox}}
\newenvironment{step2}{\begin{sBox}\begin{stp2}}{\end{stp2}\end{sBox}}

\mathtoolsset{showonlyrefs}

\begin{document}
\author[Bluhm]{Andreas Bluhm}
\email{bluhm@math.ku.dk}
\urladdr{https://orcid.org/0000-0003-4796-7633}
\address{QMATH, Department of Mathematical Sciences, University of Copenhagen, Universitetsparken 5, 2100 Copenhagen, Denmark}

\author[Capel]{Angela Capel}
\email{angela.capel@ma.tum.de}
\urladdr{https://orcid.org/0000-0001-6713-6760}
\address{Zentrum Mathematik, Technische Universit\"at M\"unchen, Boltzmannstrasse 3, 85748 Garching, Germany and Munich Center for Quantum Science and Technology (MCQST), M\"unchen, Germany}

\author[Pérez-Hernández]{Antonio Pérez-Hernández}
\email{antperez@ind.uned.es}
\urladdr{https://orcid.org/0000-0001-8600-7083}
\address{Departamento de Matem\'{a}tica Aplicada I, Escuela T\'{e}cnica Superior de Ingenieros Industriales, Universidad Nacional de Educación a Distancia, calle Juan del Rosal 12, 28040 Madrid (Ciudad Universitaria)}

\title{Weak quasi-factorization for the Belavkin-Staszewski relative entropy}

\date{\today}

\begin{abstract}
    Quasi-factorization-type inequalities for the relative entropy have recently proven to be fundamental in modern proofs of modified logarithmic Sobolev inequalities for quantum spin systems. In this paper, we show some results of weak quasi-factorization for the Belavkin-Staszewski relative entropy, i.e.\  upper bounds for the BS-entropy between two bipartite states in terms of the sum of two conditional BS-entropies, up to some multiplicative and additive factors.  
\end{abstract}

\maketitle

\section{Introduction}\label{sec:BSRS}

Given a finite-dimensional bipartite Hilbert space $\hs_{AB}=\hs_A \otimes \hs_B$ and $\rh, \sh \in \mathcal{D}(\hs_{AB})$ two density matrices on it, their \textit{(Umegaki) relative entropy} \cite{OhyaPetz-Entropy-1993} is given by
\begin{equation*}
 D(\rh \| \sh ) := \tr[\rh (\log \rh - \log \sh)] \, ,
\end{equation*}
if $\text{supp}(\rh) \subset \text{supp}(\sh)$, and by $+ \infty $ otherwise (let us assume hereafter that $\rh, \sh$ are full-rank, i.e.\ positive definite, states, so that we are always in the first case).  It constitutes the most natural extension of the Kullback-Leibler (KL) divergence \cite{KullbackLeibler-KLD-1951} to the quantum setting and provides a measure of the distinguishability between two quantum states. 

Due to the non-commuting nature of the quantum setting, the KL-divergence finds another possible quantum extension in terms of the so-called \textit{Belavkin-Staszewski relative entropy} \cite{BelavkinStaszewski-BSentropy-1982} (BS-entropy in short), which is given for two density matrices   $\rh, \sh \in \mathcal{D}(\hs_{AB})$  by
\begin{equation*}
 \bs \rh \sh := \tr[\rh \log \left( \rh^{1/2} \, \sh^{-1} \, \rh^{1/2} \right)] \, .
\end{equation*}
 In the past years, we have come to see a recent increase of interest on the BS-entropy, with a thorough study of the most fundamental properties of its generalization to maximal $f$-divergences \cite{Matsumoto2018, HiaiMosonyi-f-divergences-2017}, a strengthened data-processing inequality for such divergences \cite{BluhmCapel-BSentropy-2019} and the application of a subclass of them to estimate channel capacities \cite{FangFawzi-GeometricRenyiDivergences-2019}, among some other works. The BS-entropy is the main object of study in the current paper.

The two forms of relative entropies we discussed do not coincide in general, although the following inequality is always satisfied:
\begin{equation}\label{ineq:RE-BS}
    \ent \rh \sh \leq \bs \rh \sh \, .
\end{equation}
Moreover, the inequality is strict if, and only if, $[\rh, \sh] \neq 0$ (see e.g.\ \cite[Theorem 4.3]{HiaiMosonyi-f-divergences-2017}). Additionally, the BS-entropy agrees with the relative entropy in some of the basic properties that they satisfy, namely those of unitary invariance, additivity, continuity and the existence of a data-processing inequality, among others. 

Coming back to the Umegaki relative entropy, one of its fundamental properties is that of \textit{superadditivity}, which states that for $\rh, \sh \in \mathcal{D}(\hs_{AB})$ the following inequality holds:
\begin{equation*}
    \ent \rh {\sa \otimes \sb} \geq \ent \ra \sa + \ent \rb \sb \, .
\end{equation*}
This property  was recently extended in \cite{CapelLuciaPerezGarcia-Superadditivity-2018} to a more general setting in which the second input of the relative entropy in the LHS is not necessarily a tensor product, namely
\begin{equation}\label{eq:superadditivityRE}
    \left( 1 + 2 \norm{H(\sigma_{AB})}_\infty \right) \ent \rh \sh \geq \ent \ra \sa + \ent \rb \sb \, , 
\end{equation}
where 
\begin{equation*}
    H(\sh):= \left(\sa^{-1/2} \otimes \sb^{-1/2}\right) \sh \left(\sa^{-1/2} \otimes \sb^{-1/2}\right) - \identity_{AB} \,  
\end{equation*}
measures how far $\sh$ is from a tensor product. This inequality is an improvement to the one directly obtained from the data-processing inequality whenever $\norm{H(\sigma_{AB})}_\infty < 1/2$. In a subsequent paper \cite{CapelLuciaPerezGarcia-CRE-2018}, it was shown that Equation \eqref{eq:superadditivityRE} is equivalent to a \textit{quasi-factorization of the relative entropy}, i.e.\ 
\begin{equation}\label{eq:QF-RE}
    \ent \rh \sh \leq \frac{1}{1- 2  \norm{H(\sigma_{AB})}_\infty } \left[ \entA A \rh \sh + \entA B \rh \sh \right] \, ,
\end{equation}
where $\entA A \rh \sh$ (analogously for $B$) is called \textit{conditional relative entropy} in $A$ and is given by 
\begin{equation*}
    \entA A \rh \sh := \ent \rh \sh - \ent \rb \sb \, .
\end{equation*}
Such a result of quasi-factorization was conceived as a quantum extension of the classical results of quasi-factorization for the entropy obtained in \cite{Cesi-QuasiFactorization-2001} and \cite{DaiPraPaganoniPosta-ClassicalMLSI-2002}, which play a key role in their proofs of the existence of a positive modified logarithmic Sobolev inequality (MLSI in short) for classical spin systems. Indeed, following this motivation, a version of Equation \eqref{eq:QF-RE} for a multipartite Hilbert space and $\sigma$ a tensor product was used in \cite{CapelLuciaPerezGarcia-CRE-2018} to obtain a uniform lower bound in the MLSI for the generalized depolarizing semigroup (see also \cite{BeigiDattaRouze-ReverseHypercontractivity-2018}). Later, a more general  version of  Equation \eqref{eq:QF-RE} for $\sigma$ a quantum Markov chain  was employed in \cite{BardetCapelLuciaPerezGarciaRouze-HeatBath1DMLSI-2019} to obtain conditions for the heat-bath dynamics to have a positive MLSI. In general, a strategy to prove MLSI for quantum spin systems via quasi-factorization of the relative entropy was the main subject of study in \cite{Capel-Thesis-2019}.

The previous inequality was extended in \cite{BardetCapelRouze-ApproximateTensorization-2020} to the more general context of finite-dimensional von Neumann algebras: Let $\mathcal{M} \subset \mathcal{N}_1 , \, \mathcal{N}_2 \subset \mathcal{N}$ be von Neumann subalgebras of the algebra of linear operators acting on a finite-dimensional Hilbert space $\hs$ and let $E^{\mathcal{M}}, \, E_1, \, E_2$ be corresponding conditional expectations onto $\mathcal{M}, \, \mathcal{N}_1, \, \mathcal{N}_2$, respectively. Then, a \textit{weak approximate tensorization } for the relative entropy with parameters $c\geq 1$, $d \geq 0$ is satisfied (and denoted by AT(c,d)) if, for any $\rho \in \mathcal{D}(\mathcal{H})$, it holds that
\begin{equation}\label{eq:weak_AT}
    \ent \rho {E_*^{\mathcal{M}}(\rho)} \leq c \left(   \ent \rho {E_{1*}(\rho)}  +  \ent \rho {E_{2*}(\rho)}  \right)  + d \, ,
\end{equation}
where the maps $E_*^{\mathcal{M}}, \, E_{1*}, \, E_{2*}$ are the Hilbert-Schmidt duals of $E^{\mathcal{M}}, \, E_1, \, E_2$, respectively (see also \cite{Laracuente-QuasiFactorization-2019}, where a strong version of approximate tensorization, with $d=0$, is considered). Following the lines presented in the aforementioned papers for classical spin systems, an inequality of the form of Equation \eqref{eq:weak_AT} was the key tool in the proof of existence of a positive MLSI for commuting classical and nearest-neighbour Schmidt semigroups in \cite{CapelRouzeStilckFranca-MLSIcommuting-2020}, providing the first examples of quantum spin lattice systems satisfying this property independently of the system size.

In the current paper, we focus on the framework provided by the BS-entropy instead of the Umegaki relative entropy. More specifically, we address the problem of proving a form of quasi-factorization for the BS-entropy, namely an analogue of Equation \eqref{eq:QF-RE} for the BS-entropy. Such a result might find future applications in the fields of quantum information theory and quantum many-body systems (as discussed above with its analogue for the Umegaki relative entropy). However, note that a (strong) quasi-factorization, without an additive term, cannot hold for the BS-entropy, since it would be equivalent to the superadditivity of the BS-entropy, which is known to fail in general. We discuss this in further detail at the end of the paper. 

Therefore, following the lines of Equation \eqref{eq:weak_AT}, our main aim is to prove results of the form
\begin{equation*}
    \bs {\rho_{AB}} \sh \leq c \left[ \bsA A \rh \sh + \bsA B \rh \sh \right] + d\, ,
\end{equation*}
for $\bsA A \rh \sh$ a suitable conditional BS-entropy in $A$ and $c>0$, $d>0$, possibly depending on  $\rh, \sh$ and such that they reduce to 
\begin{equation*}
    \bs {\rho_{AB}} \sh \leq \bsA A \rh \sh + \bsA B \rh \sh \, ,
\end{equation*}
whenever the states $\rh$ and $\sh$ are ``nice" enough. The rest of the paper is devoted to the proof of two results of this form, after the introduction of a notion of conditional BS-entropy, and a discussion on the relevance and possible improvement of the result. 

\section{Main results}

Before stating the main results of this paper, we need to introduce a suitable notion of conditional BS-entropy. For that, we follow the lines of \cite{CapelLuciaPerezGarcia-CRE-2018} for the conditional relative entropy. In a bipartite system $AB$, a conditional BS-entropy in $A$ should provide the effect of the BS-entropy of two states in the global space conditioned to the value of their BS-entropy in $B$, as a possible extension of the classical definition of conditional entropy of a function. 

\begin{defi}
Let $\hs_{AB}=\hs_A \otimes \hs_B$ be a bipartite Hilbert space and $\rho_{AB}, \sigma_{AB}$ two positive definite states on it. We define the \textit{conditional BS-entropy} in  $A$ by
\begin{equation*}
    \bsA A \rh \sh := \bs \rh \sh - \bs \rb \sb \, .
\end{equation*}
\end{defi}

Now we are in position to state and prove the following result, where $\langle A,  B\rangle_\rho := \tr[A^\ast \rho^{1/2} B \rho^{1/2}]$ is the $\rho$-weighted (or KMS) inner product \cite{Kosaki-noncommLp-1984}. 

\begin{theo}\label{thm:W-QF-BS-1}
Let $\hs_{AB}=\hs_A \otimes \hs_B$ be a bipartite Hilbert space and let $\rho_{AB}, \sigma_{AB}$ be two positive definite states on it. The following inequality holds whenever $\norm{\sigma_{AB} - \sigma_A \otimes \sigma_B}_\infty \sigma_\text{min}^{-2} < d_A d_B /2$:
\begin{equation}\label{eq:weak-QF-BS-1}
\bs {\rho_{AB}} \sh \leq M(\sigma_{AB}) \left[ \bsA A \rh \sh + \bsA B \rh \sh \right] + L(\rh, \sh) \, ,
\end{equation}
where 
\begin{equation}
    M(\sh) := \frac{1}{1-\frac{2  \, \sigma_\text{min}^{-2} }{d_A d_B} \norm{\sigma_{AB} - \sigma_A \otimes \sigma_B}_\infty }  \, ,
\end{equation}
for $\sigma_\text{min}$ the minimal eigenvalue of $\sigma_{AB}$, $d_A$ and $d_B$ the dimensions of $\hs_A$ and $\hs_B$, respectively, and
\begin{equation}
    L(\rh, \sh) := M(\sh)  \left( \left\langle \sa\otimes \sb , \sa^{-1} \otimes \sb^{-1} \right\rangle_{\ra \otimes \rb} -  1 \right) \, .
\end{equation}
Note that if $\sigma_{AB}=\sa \otimes \sb$, we have $M(\sh) = 1$, and if $\rho_A^{1/2}\sigma_A^{-1/2}$ and $\rho_B^{1/2}\sigma_B^{-1/2}$ are normal (in particular, if $[\ra, \sa]= [\rb, \sb] = 0$),  then $L(\rh, \sh)=0$, obtaining thus the expected tensorization for the BS-entropy.
\end{theo}

\begin{proof}
We split the proof of this result into three steps. In the first one, we obtain an upper bound for the difference between the BS-entropy and the two conditional BS-entropies using Equation \eqref{ineq:RE-BS} and the celebrated Golden-Thompson inequality \cite{Golden-GTinequality-1965,Thompson-GTInequality-1965}.

\vspace{0.2cm}
\begin{step}
 The following inequality holds:
\begin{align}\label{eq:step1}
    \bs \rh \sh & - \bsA A \rh \sh - \bsA B \rh \sh \nonumber \\
    & \leq \log \tr[\sh \left( \ra \otimes \rb \right)^{1/2} \left( \sa \otimes \sb \right)^{-1} \left( \ra \otimes \rb \right)^{1/2} ] \, .
\end{align}
\end{step}

\vspace{0.2cm}

For the proof of this inequality, let us first note that
\begin{align*}
     \bs \rh \sh & - \bsA A \rh \sh - \bsA B \rh \sh \\
     & = - \bs \rh \sh  + \bs \ra \sa + \bs \rb \sb \\
     & = \bs {\ra \otimes \rb} {\sa \otimes \sb} - \bs \rh \sh \, .
\end{align*}
Moreover, as a consequence of Equation \eqref{ineq:RE-BS}, the following bound holds:
\begin{align*}
     \bs \rh \sh & - \bsA A \rh \sh - \bsA B \rh \sh \\
     & \leq \bs {\ra \otimes \rb} {\sa \otimes \sb} - \ent \rh \sh \\
     & = \tr[\rho_{AB} \left( -\log \rh + \underbrace{\log \sh + \log \left( \ra \otimes \rb \right)^{1/2} \left( \sa \otimes \sb \right)^{-1} \left( \ra \otimes \rb \right)^{1/2} }_{\log \Omega} \right)] \\
     & = - \ent \rh \Omega \, .
\end{align*}
Now, normalizing $\Omega$ in the previous relative entropy, using the positiveness of such entropy for density matrices and the Golden-Thompson inequality, i.e.\ the fact that for any two Hermitian matrices $X$ and $Y$ the following inequality holds
\begin{equation*}
    \tr[\operatorname{e}^{X+Y}] \leq \tr[\operatorname{e}^{X}\operatorname{e}^{Y}] \, ,
\end{equation*}
we conclude
\begin{align*}
    - \ent \rh \Omega & \leq \log \tr \Omega \\
    & \leq \log \tr[\sh \left( \ra \otimes \rb \right)^{1/2} \left( \sa \otimes \sb \right)^{-1} \left( \ra \otimes \rb \right)^{1/2} ] \, ,
    \end{align*}
    which finishes the proof of Equation \eqref{eq:step1}. 
    
    \vspace{0.2cm}
    \begin{step}
     The following inequality holds:
\begin{align}\label{eq:step2}
 &  \log \tr[\sh \left( \ra \otimes \rb \right)^{1/2} \left( \sa \otimes \sb \right)^{-1} \left( \ra \otimes \rb \right)^{1/2} ]    \nonumber \\
    & \phantom{asdasdasd} \leq  \tr[ \, \left( \sh - \sa \otimes \sb \right) \, \left[ \ra^{1/2} \left( \sa^{-1} - \ra^{-1} \right) \ra^{1/2}\right]  \otimes \left[\rb^{1/2} \left( \sb^{-1} - \rb^{-1} \right) \rb^{1/2} \right] \, ] \nonumber \\
    & \phantom{asdasdasdas} +  \left\langle \sa\otimes \sb, \sa^{-1} \otimes \sb^{-1} \right\rangle_{\ra \otimes \rb} - 1 \, .
\end{align}
    \end{step}

\vspace{0.2cm}

First, if we add and subtract $\rho_A^{-1}$, resp. $\rb^{-1}$, from $\sa^{-1}$, resp. $\sb^{-1}$, we obtain
\begin{align*}
    & \log \tr[\sh \left( \ra \otimes \rb \right)^{1/2} \left( \sa \otimes \sb \right)^{-1} \left( \ra \otimes \rb \right)^{1/2} ] \\
    & \phantom{asdad} = \log \Bigg\lbrace \underbrace{\tr[ \, \sh \left[\ra^{1/2} \left( \sa^{-1} - \ra^{-1} \right) \ra^{1/2}\right]  \otimes \left[\rb^{1/2} \left( \sb^{-1} - \rb^{-1} \right) \rb^{1/2}\right] \, ]}_{Y_{AB}} - 1   \\
   &  \phantom{asdadsadasd} + \tr[ \, \sh  \left[ \ra^{1/2}  \ra^{-1}  \ra^{1/2} \right]  \otimes \left[ \rb^{1/2}  \sb^{-1} \rb^{1/2} \right] \, ]
  + \tr[ \, \sh  \left[ \ra^{1/2}  \sa^{-1}  \ra^{1/2} \right] \otimes \left[ \rb^{1/2}  \rb^{-1} \rb^{1/2} \right] \, ] \Bigg\rbrace \\
   & \phantom{asdad} = \log \Bigg\lbrace Y_{AB} + \underbrace{\tr[  \sa \, \ra^{1/2}  \sa^{-1}  \ra^{1/2}]}_{X_A} + \underbrace{\tr[\sb \, \rb^{1/2}  \sb^{-1}  \rb^{1/2}]}_{X_B}- 1  \Bigg\rbrace \, .
\end{align*}
Now, by virtue of the inequality $\log(x+1) \leq x$, we obtain
\begin{equation*}
    \log \tr[\sh \left( \ra \otimes \rb \right)^{1/2} \left( \sa \otimes \sb \right)^{-1} \left( \ra \otimes \rb \right)^{1/2} ] \leq Y_{AB} + X_A + X_B - 2 \, .
\end{equation*}
Next, we focus on the term $Y_{AB}$, where we add and subtract $\sa \otimes \sb \text{ from } \sh$ to obtain
\begin{align*}
   &  \tr[  \, \sh  \left[ \ra^{1/2} \left( \sa^{-1} - \ra^{-1} \right) \ra^{1/2} \right] \otimes \left[ \rb^{1/2} \left( \sb^{-1} - \rb^{-1} \right) \rb^{1/2}\right] \,  ] \\
   & \phantom{asdads}= \underbrace{\tr[ \, \left( \sh - \sa \otimes \sb \right) \, \left[ \ra^{1/2} \left( \sa^{-1} - \ra^{-1} \right) \ra^{1/2} \right] \otimes \left[\rb^{1/2} \left( \sb^{-1} - \rb^{-1} \right) \rb^{1/2}\right] \, ]}_{Z_{AB}}\\
   & \phantom{asdadsas} + \tr[\, \left[ \sa \, \ra^{1/2} \left( \sa^{-1} - \ra^{-1} \right) \ra^{1/2}\right]   \otimes \left[ \sb \, \rb^{1/2} \left( \sb^{-1} - \rb^{-1} \right) \rb^{1/2}\right] \, ] \\
   & \phantom{asdads} = Z_{AB} + X_A X_B - X_A -X_B + 1 \, .
 \end{align*}
 Therefore,
 \begin{equation*}
     Y_{AB} + X_A + X_B - 2 = Z_{AB} + X_A X_B - 1 \, \, ,
 \end{equation*}
 which yields Equation \eqref{eq:step2} after noticing that
 \begin{equation*}
     X_A X_B = \left\langle \sa\otimes \sb, \sa^{-1} \otimes \sb^{-1} \right\rangle_{\ra \otimes \rb} \,.
 \end{equation*}
 
   \vspace{0.2cm}
   
   \begin{step}
     The following inequality holds:
\begin{align}\label{eq:step3}
 &   \tr[ \, \left( \sh - \sa \otimes \sb \right) \, \left[\ra^{1/2} \left( \sa^{-1} - \ra^{-1} \right) \ra^{1/2}\right]  \otimes \left[\rb^{1/2} \left( \sb^{-1} - \rb^{-1} \right) \rb^{1/2}\right] \, ]    \nonumber \\
    & \phantom{asdasdasd} \leq \frac{2 \, \sigma_\text{min}^{-2}}{d_A d_B}   \norm{\sh - \sa \otimes \sb}_\infty \bs \rh \sh \, .
\end{align}
   \end{step}

\vspace{0.2cm}

By virtue of Hölder's inequality, it is clear that
\begin{align*}
    &  \tr[ \, \left( \sh - \sa \otimes \sb \right) \left[ \ra^{1/2} \left( \sa^{-1} - \ra^{-1} \right) \ra^{1/2} \right] \otimes \left[ \rb^{1/2} \left( \sb^{-1} - \rb^{-1} \right) \rb^{1/2} \right] \, ] \\
    & \phantom{asdasda}\leq \norm{\sh - \sa \otimes \sb }_\infty \norm{\ra^{1/2} \left( \sa^{-1} - \ra^{-1} \right) \ra^{1/2} }_1 \norm{\rb^{1/2} \left( \sb^{-1} - \rb^{-1} \right) \rb^{1/2} }_1 \, .
\end{align*}
Next, for each of the last two terms on the right-hand side, the following identity holds for any invertible matrices $X$ and $Y$:
\begin{equation*}
    X^{-1} - Y^{-1} = Y^{-1}(Y-X) X^{-1} \, .
\end{equation*}
Therefore, we have
 \begin{equation*}
     \ra^{1/2} \left( \sa^{-1} - \ra^{-1} \right) \ra^{1/2} = \ra^{-1/2} \left( \ra - \sa \right) \sa^{-1} \ra^{1/2} \, ,
 \end{equation*}
 and thus
 \begin{align*}
     \norm{\ra^{1/2} \left( \sa^{-1} - \ra^{-1} \right) \ra^{1/2}}_1 & = \norm{\ra^{-1/2} \left( \ra - \sa \right) \sa^{-1} \ra^{1/2}}_1 \, \\
     & \leq \norm{\sigma_A^{-1} ( \ra- \sa) }_1 \, \\
     & \leq \norm{\sigma_A^{-1}}_\infty \norm{ \ra- \sa}_1 \, ,
 \end{align*}
where we are using \cite[Proposition IX.1.1]{Bhatia-MatrixAnalysis-1997} in the second line and Hölder's inequality in the third one (and analogously for $B$). After bounding $\norm{\sigma_A^{-1}}_\infty $ by $\sigma_\text{min}^{-1} / d_B$, we infer Equation \eqref{eq:step3} using Pinsker's inequality \cite{Pinsker-Information-1964}, the data-processing inequality and Equation \eqref{ineq:RE-BS} in the following way:
 \begin{align*}
  \norm{\ra^{1/2} \left( \sa^{-1} - \ra^{-1} \right) \ra^{1/2} }_1 \norm{\rb^{1/2} \left( \sb^{-1} - \rb^{-1} \right) \rb^{1/2} }_1 &  \leq \frac{\sigma_\text{min}^{-2}}{d_A d_B} \norm{ \ra- \sa}_1 \norm{ \rb- \sb}_1 \\
      &  \leq  \frac{\sigma_\text{min}^{-2}}{d_A d_B} \sqrt{2 \ent \ra \sa } \sqrt{2 \ent \rb \sb} \\
       &  \leq \frac{2 \, \sigma_\text{min}^{-2}}{d_A d_B}  \ent \rh \sh  \\
        &  \leq \frac{2 \, \sigma_\text{min}^{-2}}{d_A d_B}   \bs \rh \sh \, .
 \end{align*}

\noindent To finalize the proof of the theorem, note that putting the three steps together we have:
\begin{align*}
    \bs \rh \sh \leq & \, \bsA A \rh \sh + \bsA B \rh \sh  \\
    &+ \frac{2 \, \sigma_\text{min}^{-2}}{d_A d_B}  \norm{\sh - \sa \otimes \sb }_\infty   \bs \rh \sh +  \left\langle \sa\otimes \sb , \sa^{-1} \otimes \sb^{-1} \right\rangle_{\ra \otimes \rb} -  1 \, .
\end{align*}
Thus, if $ \sigma_\text{min}^{-2}  \norm{\sh - \sa \otimes \sb }_\infty   < d_A d_B/2$, we conclude
\begin{equation*}
\bs {\rho_{AB}} \sh \leq M(\sigma_{AB}) \left[ \bsA A \rh \sh + \bsA B \rh \sh \right] + L(\rh, \sh) \, ,
\end{equation*}
with
\begin{equation*}
    M(\sh) := \frac{1}{1-\frac{2 \, \sigma_\text{min}^{-2}}{d_A d_B}  \norm{\sigma_{AB} - \sigma_A \otimes \sigma_B}_\infty  }  \, ,
\end{equation*}
 and
\begin{equation*}
    L(\rh, \sh) := M(\sh)  \left( \left\langle \sa\otimes \sb , \sa^{-1} \otimes \sb^{-1} \right\rangle_{\ra \otimes \rb} -  1 \right) \, .
\end{equation*}
 Moreover, note that if $\sh= \sa \otimes \sb$, then $\norm{\sigma_{AB} - \sigma_A \otimes \sigma_B}_\infty= 0$ and thus $ M(\sh)=1$. Instead, if $ \rho_{A}^{1/2}\sigma_{A}^{-1/2}$ and $ \rho_{B}^{1/2}\sigma_{B}^{-1/2}$ are normal (in particular, if $[\ra, \sa] = [\rb, \sb]=0$), then
 \begin{align*}
     \left\langle \sa\otimes \sb , \sa^{-1} \otimes \sb^{-1}  \right\rangle_{\ra \otimes \rb} & = \tr[(\sa \ra^{1/2} \sa^{-1} \ra^{1/2})\otimes (\sb \rb^{1/2} \sb^{-1} \rb^{1/2})   ]\\ 
     & = \tr[(\sa \sa^{-1/2} \ra \sa^{-1/2})\otimes (\sb \sb^{-1/2} \rb \sb^{-1/2})   ] = \tr[ \ra \otimes \rb] =1 \, ,
 \end{align*}
 and hence, $L(\rh, \sh) =0$, yielding
 \begin{equation}\label{eq:qf-BS}
     \bs \rh {\sa \otimes \sb} \leq\bsA A  \rh {\sa \otimes \sb}  +  \bsA B  \rh {\sa \otimes \sb} \, ,
 \end{equation}
 when both conditions hold simultaneously.
\end{proof}

In the previous result, we have provided a weak quasi-factorization for the BS-entropy. The nomenclature ``weak'' 
stems from the presence of a not necessarily vanishing additive factor, as opposed to those results of (strong) quasi-factorization, in which the latter does not appear. The necessity of such a factor in the BS-entropy setting (and, thus, the impossibility of a strong quasi-factorization for the BS-entropy)  will become clear in the next section.

Next, we provide another result of weak quasi-factorization for the BS-entropy, with the advantage that the multiplicative factor is the same as for the quasi-factorization of the relative entropy in Equation \eqref{eq:QF-RE} (see \cite{CapelLuciaPerezGarcia-CRE-2018}). Let us recall that this multiplicative factor depends on the operator norm of
\begin{equation}\label{eq:H_RE}
    H(\sh) :=  \left(\sigma_A^{-1/2} \otimes \sigma_B^{-1/2} \right) \sigma_{AB} \left(\sigma_A^{-1/2} \otimes \sigma_B^{-1/2} \right) - \identity_{AB}   \,.
\end{equation}
This operator consists of the difference between a possible inversion of $\sigma_A^{-1/2} \otimes \sigma_B^{-1/2}$ with respect to $\sigma_{AB}$ and the identity in $AB$. Then, we are in position to state and prove the next result.

\begin{theo}\label{thm:W-QF-BS-2}
Let $\hs_{AB}=\hs_A \otimes \hs_B$ be a bipartite Hilbert space and let $\rho_{AB}, \sigma_{AB}$ be two positive definite states on it. The following inequality holds whenever $\norm{H(\sh)}_\infty  < 1/2$:
\begin{equation}\label{eq:weak-QF-BS-2}
\bs {\rho_{AB}} \sh \leq \widetilde{M}(\sigma_{AB}) \left[ \bsA A \rh \sh + \bsA B \rh \sh \right] + \widetilde{L}(\rh, \sh) \, ,
\end{equation}
where 
\begin{equation}
    \widetilde{M}(\sh) := \frac{1}{1-2 \norm{ H(\sh)  }_\infty }  \, ,
\end{equation}
and
\begin{equation}
    \widetilde{L}(\rh, \sh) := \frac{1+2 \norm{ H(\sh)  }_\infty  }{1-2 \norm{ H(\sh)  }_\infty}   \big( \norm{\eta_A - \rho_A}_1 \norm{\eta_B - \rho_B}_1 + \norm{\eta_A - \rho_A}_1 + \norm{\eta_B - \rho_B}_1  \big) \, , 
\end{equation}
for 
\begin{equation}
    \eta_A := \sigma_A^{1/2} \rho_A^{1/2} \sigma_A^{-1} \rho_A^{1/2} \sigma_A^{1/2} \phantom{asd} \text{and }  \phantom{asd} \eta_B := \sigma_B^{1/2} \rho_B^{1/2} \sigma_B^{-1} \rho_B^{1/2} \sigma_B^{1/2} \, .
\end{equation}
Note that if $\sigma_{AB}=\sa \otimes \sb$, we have $\widetilde{M}(\sh) = 1$, and if $\rho_A^{1/2}\sigma_A^{-1/2}$ and $\rho_B^{1/2}\sigma_B^{-1/2}$ are normal (in particular, if $[\ra, \sa]= [\rb, \sb] = 0$),  then $\widetilde{L}(\rh, \sh)=0$, obtaining thus the expected tensorization for the BS-entropy.
\end{theo}

\begin{proof}
The proof of this result is identical to that of Theorem \ref{thm:W-QF-BS-1} up to Step 3. Contrary to the former theorem, the third step of Theorem \ref{thm:W-QF-BS-2} involves both the multiplicative and additive factors.

\vspace{0.2cm}

\begin{step2}
    The following inequality holds:
\begin{align}\label{eq:step3bis}
 &   \tr[ \, \left( \sh - \sa \otimes \sb \right) \, \left[\ra^{1/2} \left( \sa^{-1} - \ra^{-1} \right) \ra^{1/2}\right]  \otimes \left[\rb^{1/2} \left( \sb^{-1} - \rb^{-1} \right) \rb^{1/2}\right] \, ]    \nonumber \\
  & \phantom{asdasdasd}  +\left\langle \sa\otimes \sb , \sa^{-1} \otimes \sb^{-1} \right\rangle_{\ra \otimes \rb} -  1 \nonumber \\
    & \phantom{asdasda} \leq 2 \norm{H(\sh)}_\infty \bs \rh \sh   \nonumber \\
    & \phantom{asdasdasd} +  \left( 1+ 2 \norm{H(\sh)}_\infty \right) \left(  \norm{\eta_A-  \rho_A }_1  +\norm{\eta_B - \rb}_1 +\norm{\eta_A - \ra}_1 \norm{\eta_B - \rb}_1    \right) \, .
\end{align}
\end{step2}

\vspace{0.2cm}

First, to simplify the expression above, let us denote
\begin{equation}\label{eq:etas}
    \eta_A := \sigma_A^{1/2} \rho_A^{1/2} \sigma_A^{-1} \rho_A^{1/2} \sigma_A^{1/2} \phantom{asd} \text{and }  \phantom{asd} \eta_B := \sigma_B^{1/2} \rho_B^{1/2} \sigma_B^{-1} \rho_B^{1/2} \sigma_B^{1/2} \, .
\end{equation}
With this, note that 
\begin{equation}\label{eq:eta_A_otimes_eta_B}
    \left\langle \sa\otimes \sb , \sa^{-1} \otimes \sb^{-1} \right\rangle_{\ra \otimes \rb} = \tr[\eta_A \otimes \eta_B] \, ,
\end{equation}
and the first trace in the left-hand side of Equation \eqref{eq:step3bis} can be rewritten as:
\begin{align*}
 &   \tr[ \, \left( \sh - \sa \otimes \sb \right) \, \left[\ra^{1/2} \left( \sa^{-1} - \ra^{-1} \right) \ra^{1/2}\right]  \otimes \left[\rb^{1/2} \left( \sb^{-1} - \rb^{-1} \right) \rb^{1/2}\right] \, ]     \\
    & \phantom{asdasda} = \tr \Big[ \, \left( \left(\sigma_A^{-1/2} \otimes \sigma_B^{-1/2} \right)   \sh \left(\sigma_A^{-1/2} \otimes \sigma_B^{-1/2} \right) - \identity_{AB} \right) \\
    &  \phantom{asdasdasda} \cdot \left[ \sigma_A^{1/2} \ra^{1/2} \left( \sa^{-1} - \ra^{-1} \right) \ra^{1/2} \sigma_A^{1/2}\right]  \otimes \left[ \sb^{1/2}\rb^{1/2} \left( \sb^{-1} - \rb^{-1} \right) \rb^{1/2} \sb^{1/2} \right] \, \Big]      \\
    & \phantom{asdasda} = \tr[ H( \sigma_{AB}) \left( \eta_A - \sigma_A \right) \otimes \left( \eta_B - \sigma_B \right)  ] \, .
\end{align*}
Now, we add and subtract $\ra$ and $\rb$ in $\left( \eta_A - \sigma_A \right) $ and $\left( \eta_B - \sigma_B \right) $, respectively, in the previous expression. Thus, we obtain
\begin{align*}
  & \tr[ H( \sigma_{AB}) \left( (\eta_A - \ra) + (\ra-  \sigma_A) \right) \otimes \left( (\eta_B - \rb) + (\rb - \sigma_B)  \right)  ]  \\
   & \phantom{asdas} = \tr[ H( \sigma_{AB}) \left(  \ra-  \sigma_A \right) \otimes \left(\rb - \sigma_B  \right) ] +  \tr[H( \sigma_{AB}) \left( \eta_A - \ra  \right) \otimes \left( \eta_B - \rb  \right)] \\
      & \, \phantom{asdas} + \tr[ H( \sigma_{AB}) \left(  \eta_A-  \rho_A \right) \otimes \left(\rb - \sigma_B  \right) ]  + \tr[ H( \sigma_{AB}) \left(  \ra-  \sigma_A \right) \otimes \left(\eta_B - \rb  \right) ] \\
       & \phantom{asdas} \leq \norm{H(\sh)}_\infty \norm{\ra-\sa}_1 \norm{\rb-\sb}_1 \\
       & \phantom{asdas} + \norm{H(\sh)}_\infty \left(  \norm{\eta_A-  \rho_A }_1 \norm{\rb-\sb}_1 +\norm{\ra-\sa}_1 \norm{\eta_B - \rb}_1 +\norm{\eta_A - \ra}_1 \norm{\eta_B - \rb}_1    \right) \, ,
\end{align*} 
where we have used Hölder's inequality repeatedly in the last inequality. We furthermore bound the last line using $\norm{\ra-\sa}_1 , \norm{\rb-\sb}_1 \leq 2$, obtaining
\begin{align*}
    & \norm{H(\sh)}_\infty \left(  \norm{\eta_A-  \rho_A }_1 \norm{\rb-\sb}_1 +\norm{\ra-\sa}_1 \norm{\eta_B - \rb}_1 +\norm{\eta_A - \ra}_1 \norm{\eta_B - \rb}_1    \right) \\
    & \phantom{asdasdadsadasd} \leq  2 \norm{H(\sh)}_\infty \left(  \norm{\eta_A-  \rho_A }_1  +\norm{\eta_B - \rb}_1 +\norm{\eta_A - \ra}_1 \norm{\eta_B - \rb}_1    \right) \, .
\end{align*}
Proceeding similarly for the right-hand side of Equation \eqref{eq:eta_A_otimes_eta_B}, after adding and subtracting $\ra$ and $\rb$ to $\eta_A$ and $\eta_B$, respectively, we have:
\begin{align*}
    \tr[(\eta_A - \ra + \ra ) \otimes (\eta_B- \rb + \rb)] & = \tr[(\eta_A - \ra) \otimes (\eta_B- \rb) ] +  \tr[\eta_A - \ra] + \tr[\eta_B - \rb] + 1 \, \\
    & \leq \norm{\eta_A - \ra}_1 \norm{\eta_B - \rb}_1 + \norm{\eta_A-  \rho_A }_1  +\norm{\eta_B - \rb}_1 + 1 \, .
\end{align*}
Therefore, putting the previous inequalities together, the following upper bound holds for the left-hand side of Equation \eqref{eq:step3bis}:
\begin{align*}
 &   \tr[ \, \left( \sh - \sa \otimes \sb \right) \, \left[\ra^{1/2} \left( \sa^{-1} - \ra^{-1} \right) \ra^{1/2}\right]  \otimes \left[\rb^{1/2} \left( \sb^{-1} - \rb^{-1} \right) \rb^{1/2}\right] \, ]    \nonumber \\
  & \phantom{asdasdasdsd}  +\left\langle \sa\otimes \sb , \sa^{-1} \otimes \sb^{-1} \right\rangle_{\ra \otimes \rb} -  1 \nonumber \\
    & \phantom{asdasdasd} \leq  \norm{H(\sh)}_\infty \norm{\ra-\sa}_1 \norm{\rb-\sb}_1 \\
       & \phantom{asdasasdasd} + \left(1+ 2 \norm{H(\sh)}_\infty \right) \left(  \norm{\eta_A-  \rho_A }_1 + \norm{\eta_B - \rb}_1 +\norm{\eta_A - \ra}_1 \norm{\eta_B - \rb}_1    \right) \, ,
\end{align*}
To conclude the proof of this step, note that we can upper bound the term in the third line above using again the data-processing inequality, Pinsker's inequality and Equation \eqref{ineq:RE-BS}, to obtain
\begin{equation*}
    \norm{H(\sh)}_\infty \norm{\ra-\sa}_1 \norm{\rb-\sb}_1  \leq 2  \norm{H(\sh)}_\infty  \bs \rh \sh \, .
\end{equation*}
Finally, putting this together with Steps 1 and 2 in the proof of Theorem \ref{thm:W-QF-BS-1}, we get
\begin{align*}
    \bs \rh \sh \leq & \, \bsA A \rh \sh + \bsA B \rh \sh  +  2  \norm{H(\sh)}_\infty  \bs \rh \sh \\
    &+ \left(1+ 2 \norm{H(\sh)}_\infty \right) \left(  \norm{\eta_A-  \rho_A }_1 + \norm{\eta_B - \rb}_1 +\norm{\eta_A - \ra}_1 \norm{\eta_B - \rb}_1    \right) \, .
\end{align*}
Therefore, if $\norm{H(\sh)}_\infty  < 1/2$, we have 
\begin{equation*}
\bs {\rho_{AB}} \sh \leq \widetilde{M}(\sigma_{AB}) \left[ \bsA A \rh \sh + \bsA B \rh \sh \right] + \widetilde{L}(\rh, \sh) \, ,
\end{equation*}
where 
\begin{equation*}
    \widetilde{M}(\sh) := \frac{1}{1-2 \norm{ H(\sh)  }_\infty }  \, ,
\end{equation*}
and
\begin{equation*}
    \widetilde{L}(\rh, \sh) := \frac{1+2 \norm{ H(\sh)  }_\infty  }{1-2 \norm{ H(\sh)  }_\infty}   \big( \norm{\eta_A - \rho_A}_1 \norm{\eta_B - \rho_B}_1 + \norm{\eta_A - \rho_A}_1 + \norm{\eta_B - \rho_B}_1  \big) \, .
\end{equation*}
 Moreover, note that if $\sh= \sa \otimes \sb$, then $\norm{H (\sh) }_\infty= 0$ and thus $ \widetilde{M}(\sh)=1$. Instead, assume $\ra^{1/2} \sa^{-1/2}$ is normal (it works analogously for $B$). Then, if follows that
 \begin{align}\label{eq:Normal_Additive_Factor}
     \norm{\eta_A - \ra}_1 & = \norm{ \, \sa^{1/2} \left( \ra^{1/2} \sa^{-1} \ra^{1/2} - \sa^{-1/2} \ra \, \sa^{-1/2}   \right) \sa^{1/2} \, }_1 \nonumber   \\
     & \leq  \norm{ \, \ra^{1/2} \sa^{-1} \ra^{1/2} - \sa^{-1/2} \ra \, \sa^{-1/2} }_\infty \\
     & = 0  \, , \nonumber
 \end{align}
 and hence, $\widetilde{L}(\rh, \sh) =0$. Hence, if both conditions hold, we get 
 \begin{equation*}
     \bs \rh {\sa \otimes \sb} \leq  \bsA A  \rh {\sa \otimes \sb}  +  \bsA B  \rh {\sa \otimes \sb} \, .
 \end{equation*}
\end{proof}

\begin{remark}
Note that since 
\begin{equation*}
    \norm{\eta_A - \ra}_1 \leq  \norm{X X^\ast - X^\ast X }_\infty
\end{equation*}
with $X = \ra^{1/2} \sa^{-1/2}$ in Equation \eqref{eq:Normal_Additive_Factor}, $\norm{\eta_A - \ra}_1$ and $\norm{\eta_B - \rb}_1$ (and hence $\widetilde{L} (\rh, \, \sh)$) provide a measure of how far $\ra^{1/2} \sa^{-1/2}$ and $\rb^{1/2} \sb^{-1/2}$, respectively, are from being normal.

Moreover, we can bound $   \norm{\eta_A - \ra}_1$ in terms of the commutator $\left[\ra^{1/2}, \, \sa^{-1/2} \right]$. Indeed, 
\begin{align*}
     \norm{\eta_A - \ra}_1 & =\\ 
     & \hspace{-10mm} =\norm{\, \sa^{1/2} \left[\ra^{1/2}, \sa^{{-}1/2}\right] \left[\sa^{{-}1/2}, \ra^{1/2}\right] \sa^{1/2} + \sa^{1/2} \left[ \ra^{1/2}, \, \sa^{-1/2} \right] \ra^{1/2} + \ra^{1/2} \left[ \sa^{-1/2}, \, \ra^{1/2} \right]  \sa^{1/2} }_1 \\
     & \hspace{-10mm} \leq  \norm{\sa^{1/2}}_2  \norm{  \left[ \ra^{1/2}, \, \sa^{-1/2} \right]  }_\infty^{2}  \norm{\sa^{1/2}}_2 +  2 \norm{\sa^{1/2}}_2  \norm{  \left[ \ra^{1/2}, \, \sa^{-1/2} \right]  }_\infty  \norm{{\ra}^{1/2}}_2 \\
     & \hspace{-10mm} \leq \norm{  \left[ \ra^{1/2}, \, \sa^{-1/2} \right]  }_\infty^{2} + 2  \norm{  \left[ \ra^{1/2}, \, \sa^{-1/2} \right]  }_\infty \, .
\end{align*}
Therefore, $\widetilde{L} (\rh, \, \sh)$ can be bounded by a term measuring the distance $\ra$, $\sa$ and $\rb$, $\sb$ are from commuting, respectively. 
\end{remark}

\section{Discussion}

Let us compare Theorem \ref{thm:W-QF-BS-1} with the analogue for the Umegaki relative entropy of Equation \eqref{eq:QF-RE}. The multiplicative factors of both quasi-factorizations are similar in spirit, since both provide a measure of how far $\sh$ is from a tensor product between $A$ and $B$. Moreover, the one appearing in Equation \eqref{eq:QF-RE} would imply that of Equation \eqref{eq:weak-QF-BS-1} for all states commuting if we had not bounded $\norm{\sa^{-1}}_\infty$ and $\norm{\sb^{-1}}_\infty$ by $\sigma_{\text{min}}^{-1}/d_B$ and $\sigma_{\text{min}}^{-1}/d_A$ to simplify the final expression. 

A clear difference between both results appears in the fact that, in our results, the assumption $[\ra, \sa] = [\rb, \sb]=0$ seems to be necessary to get back the usual tensorization for the BS-entropy, i.e.\ Equation \eqref{eq:qf-BS}. This appears already after the use of Golden-Thompson inequality during the proof, since for the remaining term to vanish, that assumption needs to hold, along with that of $\sh$ being a tensor product. This is different from the case for the relative entropy, in which only the latter assumption was necessary.

Furthermore, note that $[\ra, \sa] = [\rb, \sb]=0$ implies that the BS-entropies in each of the systems $A$ and $B$ coincide with the relative entropy between the same states. Thus, by assuming this and $\sh= \sa \otimes \sb$, we get that the quasi-factorization for the relative entropy implies that for the BS-entropy.

The comparison between Theorem \ref{thm:W-QF-BS-2} and equation \eqref{eq:QF-RE} yields a similar conclusion. The multiplicative factors of both results coincide completely in this case. Thus, the main difference is the presence of the additive factor $\widetilde{L}(\rh, \, \sh)$ in the theorem and the associated
additional requirement  ($\ra^{1/2} \sa^{-1/2}$ and $\rb^{1/2} \sb^{-1/2}$ being normal) for the reduction to the usual tensorization for the BS-entropy. 

\begin{center}
\begin{figure}[h]
    \centering
    \includegraphics[scale=0.35]{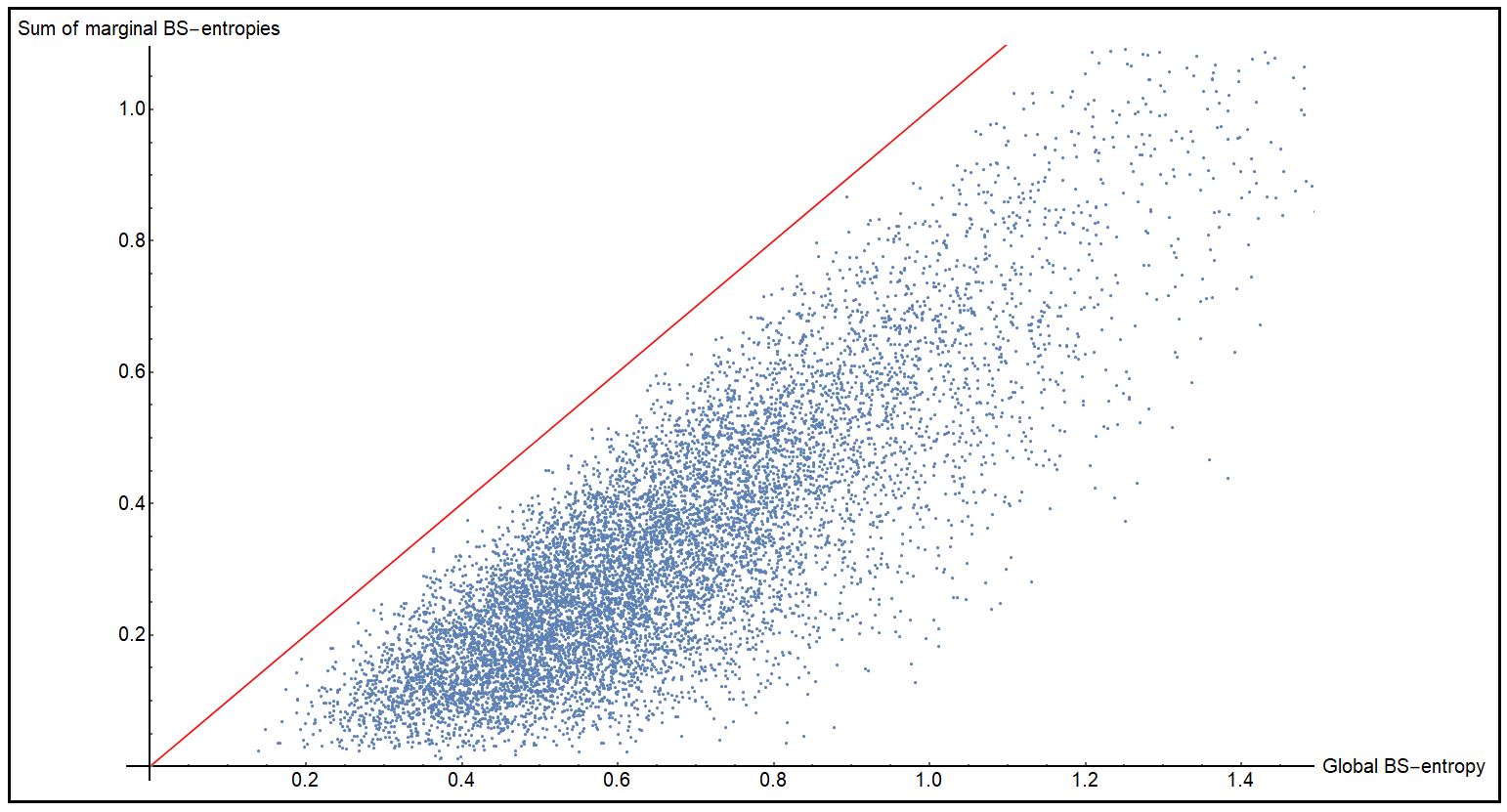} \\
    \vspace{0.3cm}  \includegraphics[scale=0.35]{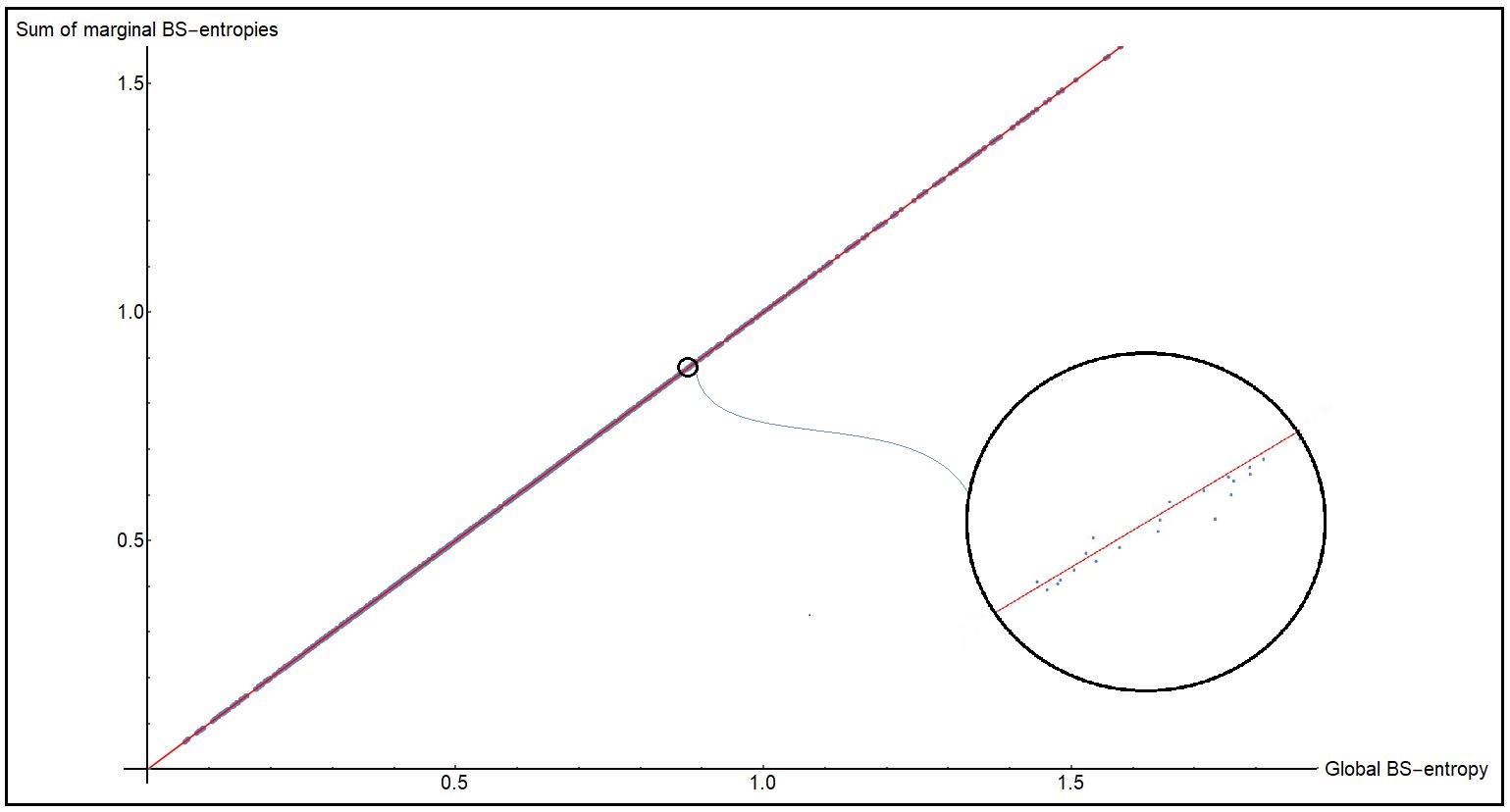} 
    \caption{Counterexample for Equation \eqref{eq:superadditivity-BS}. We represent the sum of the two conditional BS-entropies against the global BS entropy. The sample consists of $10000$ random density matrices of size $2 \times 2$. On the top, we represent the general case, whereas on the bottom we assume that the first state is a perturbation of a tensor product.}
    \label{fig:1}
\end{figure}
\end{center}

However, although the necessity for more assumptions than in the case of the relative entropy might open some discussion on the suitability of the method, no alternative would provide a quasi-factorization in the spirit of Equation \eqref{eq:weak-QF-BS-1} which reduces to Equation \eqref{eq:qf-BS} by only assuming the tensor product property on $\sh$. Indeed, by writing explicitly the conditional BS-entropies, it is easy to notice that Equation \eqref{eq:qf-BS} is equivalent to 
\begin{equation}\label{eq:superadditivity-BS}
     \bs \rh {\sa \otimes \sb} \geq  \bs \ra {\sa} + \bs \rb {\sb} \, ,
\end{equation}
and, more specifically, to the property of superadditivity for the BS-entropy. However, this property is known to be false in general for this entropy, since the properties of continuity, additivity and data processing hold for the BS-entropy and the four of them jointly provide an axiomatic characterization for the relative entropy \cite{WilmingGallegoEisert-RelativeEntropy-2017, Matsumoto-RelativeEntropy-2010}. This is also the reason for the additive factors $L(\rh, \sh)$ and $\widetilde{L}(\rh, \sh)$  being necessary in Theorems \ref{thm:W-QF-BS-1} and \ref{thm:W-QF-BS-2} respectively. Indeed, as $M(\sigma_{AB})$ reduces to $1$ whenever $\sh$ is a tensor product, if there was no additive factor then Equation \eqref{eq:weak-QF-BS-1} would reduce to Equation \eqref{eq:qf-BS} and thus be equivalent to Equation \eqref{eq:superadditivity-BS} by only assuming $\sh= \sa \otimes \sb$, reaching again a contradiction.

Additionally, a simple numerical study shows the falseness of Equation \eqref{eq:superadditivity-BS}. In Figure \ref{fig:1}, we present two graphics obtained by representing $\bs \ra {\sa} + \bs \rb {\sb}$ against $\bs \rh {\sa \otimes \sb}$ for $10000$ random density matrices of size $2 \times 2$. In the top graphic, we assume nothing more on both states, and it seems that the property of superadditivity might hold, since all the points lie below the diagonal. However, in the bottom graphic we assume that the first state is a perturbation of a tensor product, i.e.\
\begin{equation*}
    \rh := \frac{\eta_A \otimes \eta_B + \varepsilon \lambda_{AB}}{\tr[\eta_A \otimes \eta_B + \varepsilon \lambda_{AB}]} \, ,
\end{equation*}
for $\eta_{AB}, \lambda_{AB} \in \mathcal{D}(\hs_A \otimes \hs_B)$ and $\varepsilon= 0.01$. In this case, all the points are very close to the diagonal (where they would all lie for $\varepsilon=0$), but approximately 10 $\%$ of them fall above it, contradicting thus the property of superadditivity.

\vspace{0.2cm}

\noindent {\it Acknowledgments.}  AB acknowledges support from the VILLUM FONDEN via the QMATH Centre of Excellence (Grant no.  10059) and from the QuantERA ERA-NET Cofund in Quantum Technologies implemented within the European Union’s Horizon 2020 Programme (QuantAlgo project) via the Innovation Fund Denmark. AC is partially supported by a MCQST Distinguished PostDoc fellowship and by the Deutsche Forschungsgemeinschaft (DFG, German Research Foundation) under Germany's Excellence Strategy EXC-2111 390814868. APH acknowledges support from the “Juan de la Cierva Formación” fellowship (FJC2018-036519-I).

\bibliographystyle{alpha}
\bibliography{petzlit}

\vspace{0.5cm}
\end{document}